\documentclass{aastex63}
\usepackage{textcomp}

\usepackage{graphicx}
\usepackage{subfigure}
\usepackage{epsfig}
\usepackage{rotating}
\usepackage{floatrow}
\DeclareFloatFont{scriptsize}{\scriptsize}
\usepackage{amssymb}

\begin{document}

\title{An X-Ray-dim ``Isolated'' Neutron Star in a Binary?
}%

\correspondingauthor{Renxin Xu}
\email{r.x.xu@pku.edu.cn}
\correspondingauthor{Jianrong Shi}
\email{sjr@nao.cas.cn}
\author{Jie Lin}
\affiliation{Department of Astronomy, Peking University, Beijing 100871, People's Republic of China}
\affiliation{State Key Laboratory of Nuclear Physics and Technology, School of Physics, Peking University, Beijing 100871, People's Republic of China}
\affiliation{Kavli Institute for Astronomy and Astrophysics, Peking University, Beijing 100871, People's Republic of  China}

\author{Chunqian Li}
\affiliation{CAS Key Laboratory of Optical Astronomy, National Astronomical Observatories, Chinese Academy of Sciences, Beijing 100101, People's Republic of China}
\affiliation{School of Astronomy and Space Science, University of Chinese Academy of Sciences, Beijing 100049, People's Republic of China}

\author{Weiyang Wang}
\affiliation{Department of Astronomy, Peking University, Beijing 100871, People's Republic of China}
\affiliation{State Key Laboratory of Nuclear Physics and Technology, School of Physics, Peking University, Beijing 100871, People's Republic of  China}
\affiliation{Kavli Institute for Astronomy and Astrophysics, Peking University, Beijing 100871, People's Republic of  China}

\author{Heng Xu}
\author{Jinchen Jiang}
\affiliation{National Astronomical Observatories, Chinese Academy of Sciences, Beijing 100101, People's Republic of China}

\author{Daoye Yang}
\affiliation{CAS Key Laboratory of Optical Astronomy, National Astronomical Observatories, Chinese Academy of Sciences, Beijing 100101, People's Republic of China}

\author{Shahidin Yaqup }
\author{Abdusamatjan Iskanda}
\author{Shuguo Ma}
\author{Hubiao Niu}
\author{Ali Esamdin}
\affiliation{XinJiang Astronomical Observatory, Chinese Academy of Sciences}

\author{Shuai Liu}
\affiliation{CAS Key Laboratory of Optical Astronomy, National Astronomical Observatories, Chinese Academy of Sciences, Beijing 100101, People's Republic of China}

\author{Gavin Ramsay}
\affiliation{Armagh Observatory and Planetarium, College Hill, Armagh BT61 9DG, UK}

\author{Jose I. Vines}
\affiliation{Departamento de Astronom´ıa, Universidad de Chile, Casilla 36-D, Santiago, Chile}


\author{Jianrong Shi}
\affiliation{CAS Key Laboratory of Optical Astronomy, National Astronomical Observatories, Chinese Academy of Sciences, Beijing 100101, People's Republic of China}
\affiliation{School of Astronomy and Space Science, University of Chinese Academy of Sciences, Beijing 100049, People's Republic of China}

\author{Renxin Xu}
\affiliation{Department of Astronomy, Peking University, Beijing 100871, People's Republic of China}
\affiliation{State Key Laboratory of Nuclear Physics and Technology, School of Physics, Peking University, Beijing 100871, People's Republic of China}
\affiliation{Kavli Institute for Astronomy and Astrophysics, Peking University, Beijing 100871, People's Republic of  China}

\begin{abstract}
We report the discovery of a dark companion to 2MASS\,J15274848+3536572 with an orbital period of 6.14 hr. Combining the radial velocity from LAMOST observations and modelling of the multiband light curve, one obtains a mass function of $\simeq 0.131~\rm M_{\odot}$, an inclination of $45.20^\circ{}^{+0.13^{\circ}}_{-0.20^{\circ}}$, and a mass ratio of $0.631^{+0.014}_{-0.003}$, which demonstrate the binary nature of the dark companion with mass of $0.98 \pm 0.03\rm M_{\odot}$ and a main-sequence K9-M0 star of $0.62 \pm 0.01~\rm M_{\odot}$. LAMOST optical spectra at a range of orbital phases reveal extra-peaked $\rm H_{\alpha}$ emission that suggests the presence of an accretion disk. The dark companion does not seem to be a white dwarf because of the lack of any observed dwarf nova outbursts in the long-term data archive, although a magnetic white dwarf cannot be excluded. Alternatively, we propose a scenario wherein the dark companion is a neutron star, but we have not detected radio pulsations or a single pulse from the system with the FAST (Five-hundred-meter Aperture Spherical radio Telescope), which hints at a radio-quiet compact object. If the dark companion is identified as a neutron star, it will be the nearest ( $ \sim 118$\,pc) and lightest neutron star. Furthermore, a kinematic analysis of the system's orbit in the galaxy may suggest its supernova event is associated with the radionuclide $^{60} \rm Fe$ signal observed from the deep-sea crusts.
This radio-quiet and X-ray-dim nearby neutron star may resemble an XDINS (X-ray-dim isolated neutron star), but in a binary.
\end{abstract}


\section{Introduction}

The puzzling equation of the state of dense matter inside neutron stars (NSs) is focused on both microphysics and astrophysics, and moreover, the diversity of NS manifestations challenges a simple belief about their astronomical origins~\citep[e.g.,][]{2010PNAS..107.7147K}.
Among the populations in the NS zoo, there is a small group called X-ray-dim isolated neutron stars (XDINSs), characterized by low thermal X-ray luminosity within a few hundreds of parsecs.
Is there any XDINS-like compact object in a binary? How can one find it?
This is the focus of the present work. 

Up to now, only seven XDINSs have been discovered by the ROSAT all-sky observations, which are nicknamed as the Magnificent Seven \citep{Voges96}. They have an intriguing feature, that is, they are characterized by Planck-like spectra in X-ray bands as well as in optical bands but with an ordered excess, and by the nondetection of radio signals \citep{Haberl07}.
XDINSs are peculiar objects for revealing the equation of state (EOS) at supranuclear density and offer an unprecedented opportunity to study atmospheric emission on the surface (e.g., \citealt{Ho07,Wang17,Wang18}).
If similar objects can be discovered outside our local volume, it will be significant for understanding their properties as a group and their relationship to other galactic isolated NS (INS) families.
A promising INS candidate was found from the 2XMMp catalog of serendipitous sources \citep{Pires09}, and then four newly discovered INS (isolated neutron star) candidates were reported from the 4XMM-DR10 catalogue \citep{Rigoselli22}.
Such stars are measured to be close to us ($\sim120-500$\,pc), and as numerous as young radio and $\gamma$-ray pulsars locally, suggesting that there may be lots of similar Galactic sources still unknown.

%
We focus on the binary systems with compact objects, which can provide an opportunity to develop the accretion disk model 
and test the prediction of binary interaction theories \citep{2017PASP..129f2001M,2012ARA&A..50..107L}.
Interestingly, NSs with the XDINS-like characteristics, which are of great interest in the EOS problem, have not been found in binaries yet (e.g., \citealt{Pires15}).
In addition, the discovery and observed mass distributions of NSs and black holes (BHs) in the Milky Way is crucial for understanding core-collapse supernovae and the evolution of massive stars \citep{2012ApJ...749...91F,2020ApJ...896...56W}. So far, most compact binaries with NSs, or BHs are identified 
from the radio, X-ray, and gamma-ray surveys and from the gravitational waves. However, these compact binaries based on above methods may represent only a small fraction of the overall population, resulting from accreting at a sufficiently low rate or in long quiescent periods. Here, radial-velocity modulation in optical spectra is a promising method to uncover unseen compact objects that have a stellar companion \citep{2019ApJ...872L..20G,2019Natur.575..618L,2019Sci...366..637T,2022ApJ...940..165Y}. More recently, a nonaccreting neutron star candidate was discovered by synergizing optical time-domain spectroscopy \citep{2022NatAs...6.1203Y}.                     

Here we present the first optical spectroscopy and multiband optical photometry of a single-line binary, 2MASS J15274848+3536572 (hereafter J1527), and report that it consists of a K9-M0 main-sequence and an unseen companion of $ M_{c}=0.98 \pm 0.03$\,$\rm M_{\odot}$. We also discuss two possible scenarios wherein the dark companion of J1527 is a white dwarf or NS candidate, respectively. This paper is organized as follows. We describe observations of the source in Section 2 and analyze these observations to extract the physical parameters of the source in Section 3. A general discussion of the dark companion in Section 4. Our conclusions are presented in Section 5.

\section{Observations}

\subsection{Optical photometry and open-source data}

J1527 was also observed with the Nanshan 1m Wide-field Telescope (NOWT) \citep{2020RAA....20..211B} of Xinjiang Astronomical Observatory  from 2022 April 15 to 17 for three nights in the B, V and R bands. The exposure time for each filter is 30s, and the total observation time of each run in B, V, and R bands is about 220s. All the raw images for the three bands were overscan-corrected, bias-subtracted, and flat-fielded using IRAF \citep{1986SPIE..627..733T,1993ASPC...52..173T}. The optimized aperture photometry was employed using the software Sextractor \citep{1996A&AS..117..393B}. The resulting differential photometry is relative to a nearby stable reference star in the field, which is insensitive to thin clouds or moderate atmospheric variability. 

J1527 was classified as a contact eclipsing binary in the All-Sky Automated Survey (ASAS) and the Catalina Sky surveys (CSS) with a period of$\rm P=0.2556686$\,d \citep{2014ApJS..213....9D,2019MNRAS.486.1907J}. We obtained sampled light curves from ASAS and CSS and a densely sampled light curve from the SuperWASP with $\sim$ 300 s cadence and the Transiting Exoplanet Survey Telescope (TESS) and Hungarian Automated Telescope with $\sim$ 300 s cadence (HAT )\citep{2011AJ....141..166H}, which is shown in Figure 1. TIC 16320250 \footnote{The data are available at the Mikulski Archive for Space Telescopes (MAST) at the Space Telescope Science Institute: \dataset[10.17909/e21w-k824]{\doi{10.17909/e21w-k824}}} was observed by TESS in Sector 24 from 2020 April 16 to 2020 May 12. TIC 16320250 was observed with 30 minute cadence.

In addition, it also was observed with GALEX as part of the Medium Imaging Survey with the measured fluxes of (2.42 $\pm$ 0.34)\,$\times10^{-13}$ and (1.25 $\pm$ 0.05)\,$\times10^{-12}$$\rm erg$ $\rm cm^{-2}s^{-1}$ in the far- and near-ultraviolet (FUV and NUV), respectively. In {\it Gaia} EDR3, the source$\_$id of J\,1527 is 1375051479376039040 \citep{2021A&A...649A...1G}. Its EDR3 parallax of $\omega_{\rm EDR3}=8.44844 \pm 0.011068$\,mas implies a distance of d=118.13 $\pm 0.86$\,pc, which is consistent with its Gaia DR2 parallax. Gaia DR2 also reports a luminosity $L_{\rm Gaia}=0.101161 \pm 0.000605\rm~L_{\odot}$, temperature $T_{\rm eff,Gaia}=4047_{-109}^{+171}$\,K, and radius $R_{\rm Gaia}=0.65_{-0.05}^{+0.03}~R_{\odot}$ for the star \citep{2018A&A...616A...1G}.

\begin{figure}[htp]
\centering
\includegraphics[width=15cm]{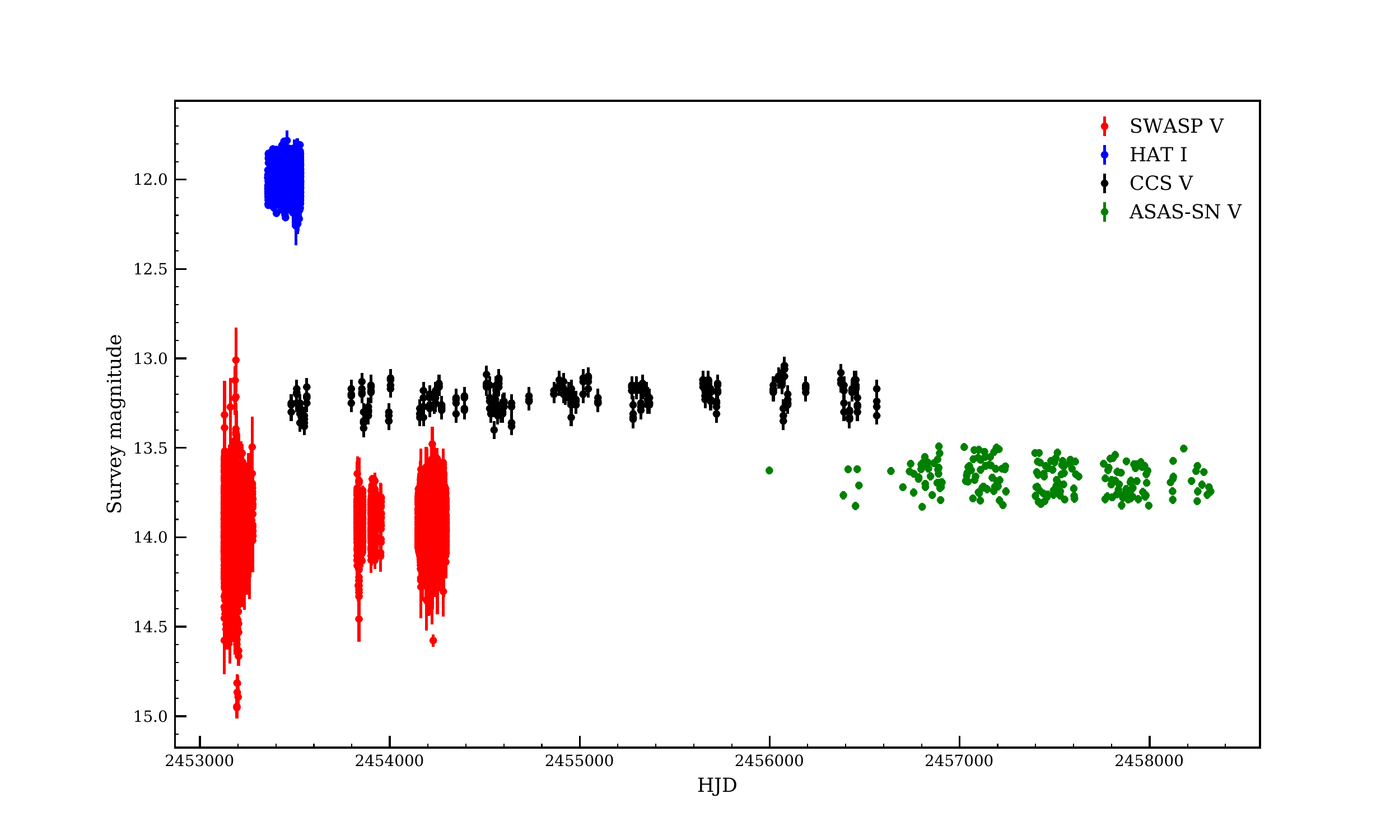}
\caption{SuperWASP, HATNet, CCS, and ASAS-SN photometry for J1527 from 2004 May 2 to 2018 July 22. No evidence for outbursts appears in the total 5193\,day time span.}
\end{figure}

\subsection{Optical spectra}
The Large Sky Area Multi-Object fiber Spectroscopic Telescope (LAMOST) is a 4m Schmidt telescope with a $5^ \circ$ field of view and has become the first spectroscopic survey to collect tens of millions of spectra from the universe with both the medium-resolution and low-resolution modes. The medium resolution observations have R$\sim$7500 with both a blue and a red arms at a limiting magnitude $G=$15 mag \citep{2020ApJS..251...15Z}. The wavelength coverage of the blue arm is from 495\,nm to 535\,nm, while it is from 630\,nm to 680\,nm for the red arm. The low-resolution observation mode has a limiting magnitude as faint as r$\sim$17.8\,mag. with a resolution R $\sim$1800, and the wavelength  range is from 370\,nm to 900\,nm \citep{2012RAA....12..723Z}.

Here, we obtained 2 low-resolution spectra and 10 medium-resolution spectra of J1527 from the LAMOST archive data. The low-resolution LAMOST spectra indicate it to be a K9-M0 type star (Figure 2) with clear signs of the \ion{Ca}{2}\,H\&K and Balmer $\rm H_{\alpha}$ and $\rm H_{\beta}$ emission lines, which tell us it has chromospheric activity. However, the $\rm H_{\alpha}$ emission shows wider range and a complex morphology including an extra peak emission profile in many observations, which may suggest the presence of an accretion disk.  We derived the barycentric velocity of each spectrum through the cross-correlation technique. Here, the emission lines of two low-resolution spectra have been excised in the cross-correlation process. For 10 medium-resolution spectra, only the spectra of the blue arms has been chosen to measure radial velocities due to the lack of emission lines in the blue range. The cross-correlation has been performed using the spectral template ($T_{\rm eff}=4500 \rm K$, logg=4.5, [Fe/H]=-1.0) calculated with the Kurucz model \citep{2003IAUS..210P.A20C}. 

In addition,  we estimate the projected rotational velocity $v\rm sin$$i$ using the red arms of the medium-resolution LAMOST spectra because of its high signal-to-noise ratio.  A $\mathcal{X}^2$ minimization was performed to constrain the projected rotational velocity $v\rm sin$$i$ using the above spectral template  with a range of $v\rm sin$$i$ from 0 to 300 km/s. The mean value obtained from the six medium-resolution LAMOST spectra is $v\rm sin$$i$$=94\pm 5 \rm km/s$. The quoted uncertainty in this value is the standard deviation. Furthermore, high-resolution spectra will confirm and improve this measurement in the future.

\begin{figure}[htp]
\centering
\includegraphics[width=15cm]{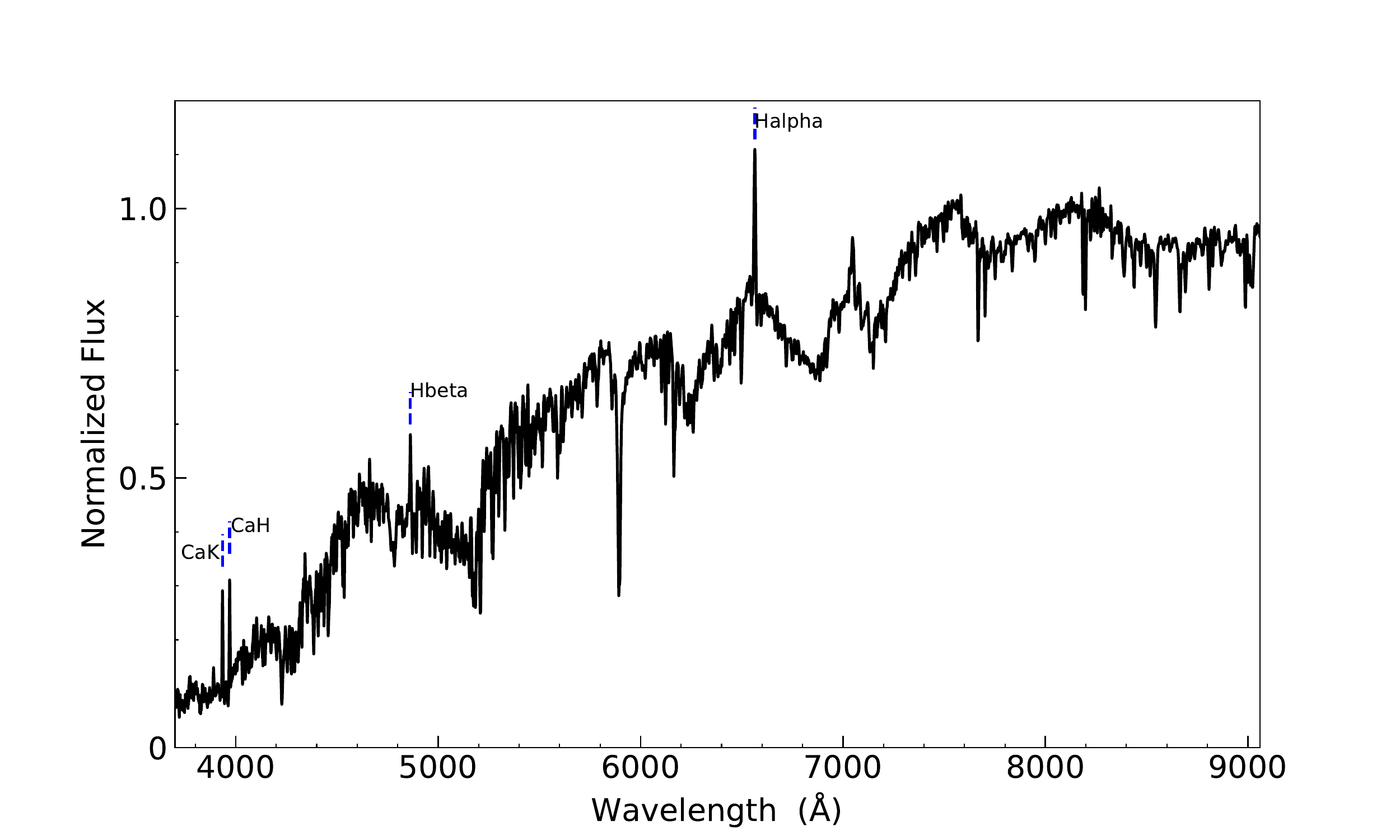}
\caption{Low-resolution LAMOST spectrum from 2020 January 22, showing the K9-M0 spectral type star and the clear \ion{Ca}{2} H\&K, Balmer $\rm H_{\alpha}$, and $\rm H_{\beta}$ emission lines.}
\end{figure}

\subsection{Radio and X-Ray observations}
We searched for radio pulsations using the FAST (Five-hundred-meter Aperture Spherical radio Telescope). The central frequency and bandwidth of the receiver of FAST are 1250 and 400 MHz, respectively \citep{2020RAA....20...64J}. Observations were acquired using the central beam of the 19 beam receiver, which is 25 K for the system temperature. J\,1527 was observed for 40 minutes and 50 minutes on 2021 October 28 and 2022 September 14, respectively. Using the PRESTO pulsar search suite \citep{2001PhDT.......123R}, we excised radio frequency interference and performed an acceleration search to retain sensitivity by setting the zmax value to be 200, as well as trial dispersion measures between 0 and 22 pc $\rm cm^{-3}$. In addition, 1RXS J152748.8+353658 was listed in the ROSAT ALL-Sky Survey Source Catalogue as a faint X-ray source \citep{1999A&A...349..389V}, with a count rate of 0.0316\,cts\,$\rm s^{-1}$ in the 0.1-2.4\,keV PSPC range. More recently, the RASS was reprocessed by \cite{2016A&A...588A.103B}. The count rate (count rate 0.0342 $\pm$ 0.0122\,cts\,$\rm s^{-1}$) and the hardness ratio (HR1\footnote{The definitions of the HR1 and HR2 are shown in Appendix A of \citep{2016A&A...588A.103B}}=0.77$\pm$0.27;HR2=-0.66$\pm$0.19) were updated, which indicate a relatively soft emission spectrum. The observed flux is 5.54$\pm$2.98\,$\times10^{-14}$$\rm erg$\,$\rm cm^{-2}s^{-1}$ (0.1-2.4\,keV) using the energy to count conversion factor (ECF=1.62$\times10^{-12}$$\rm\,erg$\,$\rm cm^{-2}s^{-1}/counts$) adopting a
blackbody model with a temperature of $ \sim 95.6$ eV and a column density of $ \sim 0.41\times 10^{20}\rm cm ^{-2}$ as in \cite{2004A&A...424..635H}.

\section{Results}
\subsection{Properties of the main-sequence K9-M0 star}
We obtained the surface temperatures ($T_{\rm eff}$) and radius ($R$) of the K9-M0 type star by fitting the spectral energy distribution (SED) using the astroARIADNE \citep{2022MNRAS.513.2719V}. For the SED, we used the photometry from {\it Gaia} DR3 ($G$, $G_{\rm BP}$, and $G_{\rm RP}$), 2MASS (J, H and K$_{\rm s}$), SDSS ($u$, $g$, $r$, $i$ and $z$), Pan-STARRS ($g$ and $y$), JOHNSON (B and V), WISE (W1 and W2) and TESS(T). We found an excellent fit by using the extinction parameters $A_{\rm v}=0$ derived from three-dimensional dust maps of \cite{2019ApJ...887...93G}, with $T_{\rm eff}=3896.42_{-46.12}^{+32.70}$\,K and $R=0.689_{-0.012}^{+0.020}~R_{\odot}$ (Figure 3), which are consistent with that of {\it Gaia} DR2. However, the parameter extinction (0.38) from DR2 in the G-band is different from that in the three-dimensional dust maps of \cite{2019ApJ...887...93G}. This may result from \cite{2019ApJ...887...93G} leveraging a greater number of photometric passbands compared to two independent passbands for Gaia DR2 and delivering typical
reddening uncertainties that are lower. In addition, the SED fit yields $T_{\rm eff}=3919.02_{-39.58}^{+30.55}$\,K, $R=0.685_{-0.010}^{+0.016}~R_{\odot}$ and $A_{\rm v}=0.03_{-0.02}^{+0.01}$ using the free extinction parameters. The results of the SED suggest the spectral type of the main-sequence star is K9-M0 and demonstrate the presence of a UV excess in the two GALEX measurements. 

Figure~4 shows the variations of the $\rm H_{\alpha}$ emission-line profiles with orbital phase, which are corrected to the rest frame of the K9-M0 star and companion, respectively. We found that the $\rm H_{\alpha}$ profiles almost peak directly at or very near the velocity of the K9-M0 star (left panel in Figure 4). This suggests that the $\rm H_{\alpha}$ emission results from the visible star's surface. The $\rm H_{\alpha}$ emission tracks the orbital motion in the rest frame of the K9-M0 type star. In addition, we see the \ion{Ca}{2}\,H\&K emission in the low-resolution spectra, which suggested that the $\rm H_{\alpha}$ emission could be from chromospheric activity as discussed in Section 2.2. The typical full width at half-maximum (FWHM) of the $\rm H_{\alpha}$ and $\rm H_{\beta}$ emission profile is $\sim 290 \rm km s^{-1}$ at  most cases, which may suggest $\rm H_{\beta}$ also results from chromospheric activity.

In addition, the $\rm H_{\alpha}$ emission profile occasionally shows a broad and narrow morphology at the specific phases ($\phi=0.194;0.743;0.805$). Here, we transfer the $\rm H_{\alpha}$ emission to the rest frame of the unseen object using the binary mass ratio, systemic velocity, and the radial velocity of the K9-M0-type star (Right in Figure 4). We find that the peak of the narrow $\rm H_{\alpha}$ component  is almost in phase with the unseen companion at the specific phases. Furthermore, in order to confirm  whether the narrow component is from an accretion disk, we measure equivalent widths (EW) and radial velocities for a narrow and broad $\rm H_{\alpha}$ at the specific phases, respectively. The median EW of the narrow component (FWHM $\sim 125 \rm km s^{-1}$) and broad component (FWHM $\sim 250 \rm km s^{-1}$) is EW$(\rm H_{\alpha,N}=0.53\pm0.13)$ and EW$(\rm H_{\alpha,B}=2.33\pm0.14)$, respectively. We find that the narrow component is indeed in antiphase with the radial-velocity curve of the visible star (See Figure 5), while the broad component is in phase with the absorption line radial velocity curve. Although the presence of a broad, double-peaked $\rm H_{\alpha}$ emission profile has generally been accepted as evidence for an accretion disk in the system \citep{2009ApJ...703.2017W,2015ApJ...804L..12S}, the shape of the $\rm H_{\alpha}$ emission lines from an accretion disk depends on an inclination angle and the choice of power-law n and disk base density
$\rho_{0}$ \citep{2010ApJS..187..228S}. Furthermore, the narrow $\rm H_{\alpha}$ component from an accretion disk may imply a small accretion disk around the compact object and/or a level of accretion disk activity.

One reasonable explanation for complex $\rm H_{\alpha}$ emission may arise from a combination of an accretion disc around the dark companion and the stellar chromosphere. Here, the narrow extra-peaked $\rm H_{\alpha}$ emission profile from an accretion disk is hard to identify when the chromosphere of the star is the main contributor to the emission. However, we also note that the $\rm H_{\alpha}$ component from an accretion disk generally has an FWHM of the order of several hundreds to a thousand km/s. Overall, we interpret the structured peak of the $\rm H_{\alpha}$ emission as evidence of an accretion disk, with complicated phenomenology that deserves further study. The high-resolution and adequate phase-coverage spectroscopic observations will help us to uncover on their true origin.

\begin{figure}[htp]
\centering
\includegraphics[width=15cm]{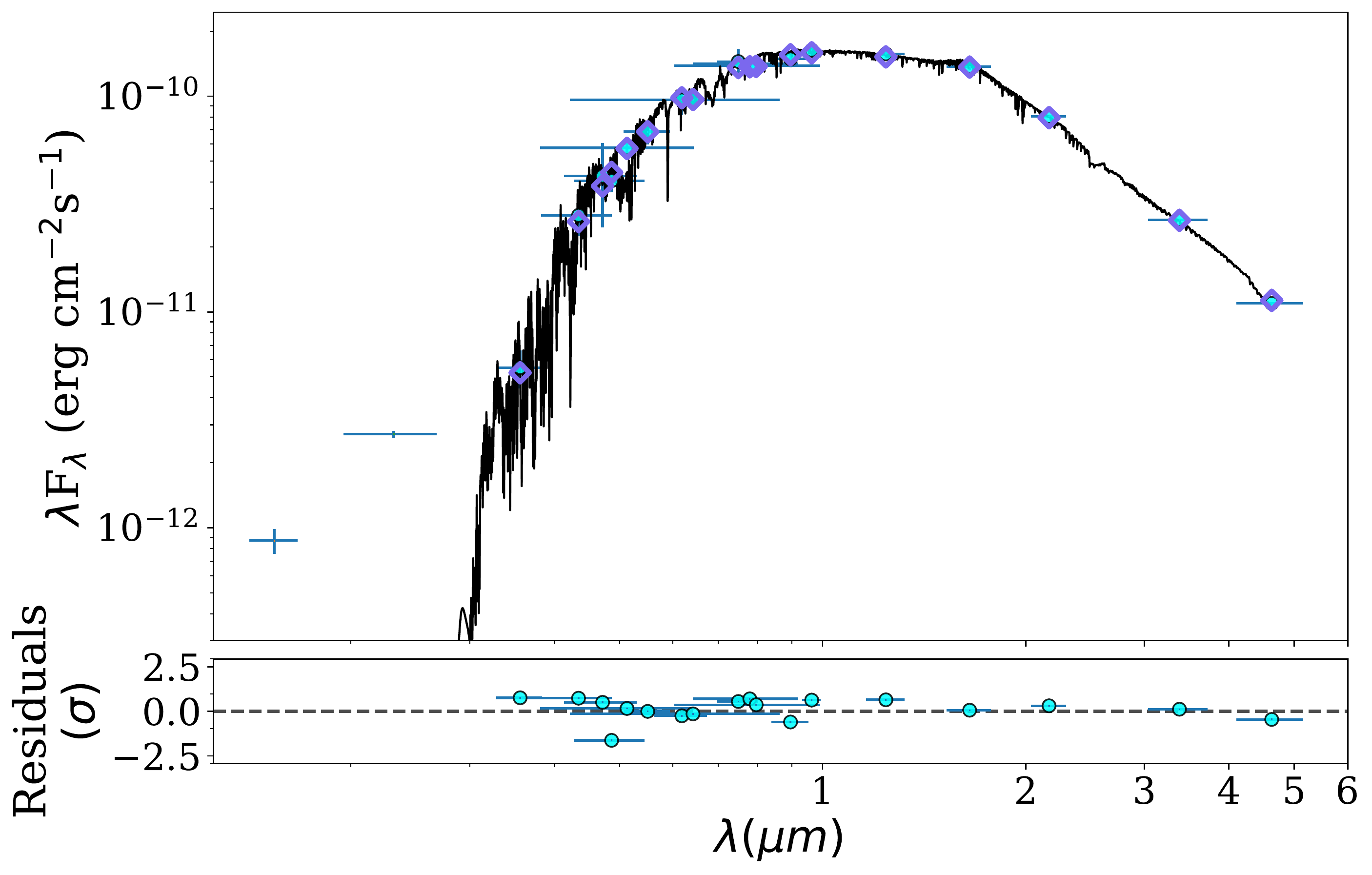}
\caption{The best-fitting SED model for J1527 after fixing the extinction parameters $A_{\rm v}=0$. The black curve is the best-fitting model. The green pluses and circles are the retrieved photometric measures. The blue diamonds are synthetic photometry. Here, the GALEX FUV and NUV (two green pluses) were not included in the SED fitting.}
\end{figure}
\begin{figure}
  \centering
\includegraphics[angle=0,width=0.49\textwidth]{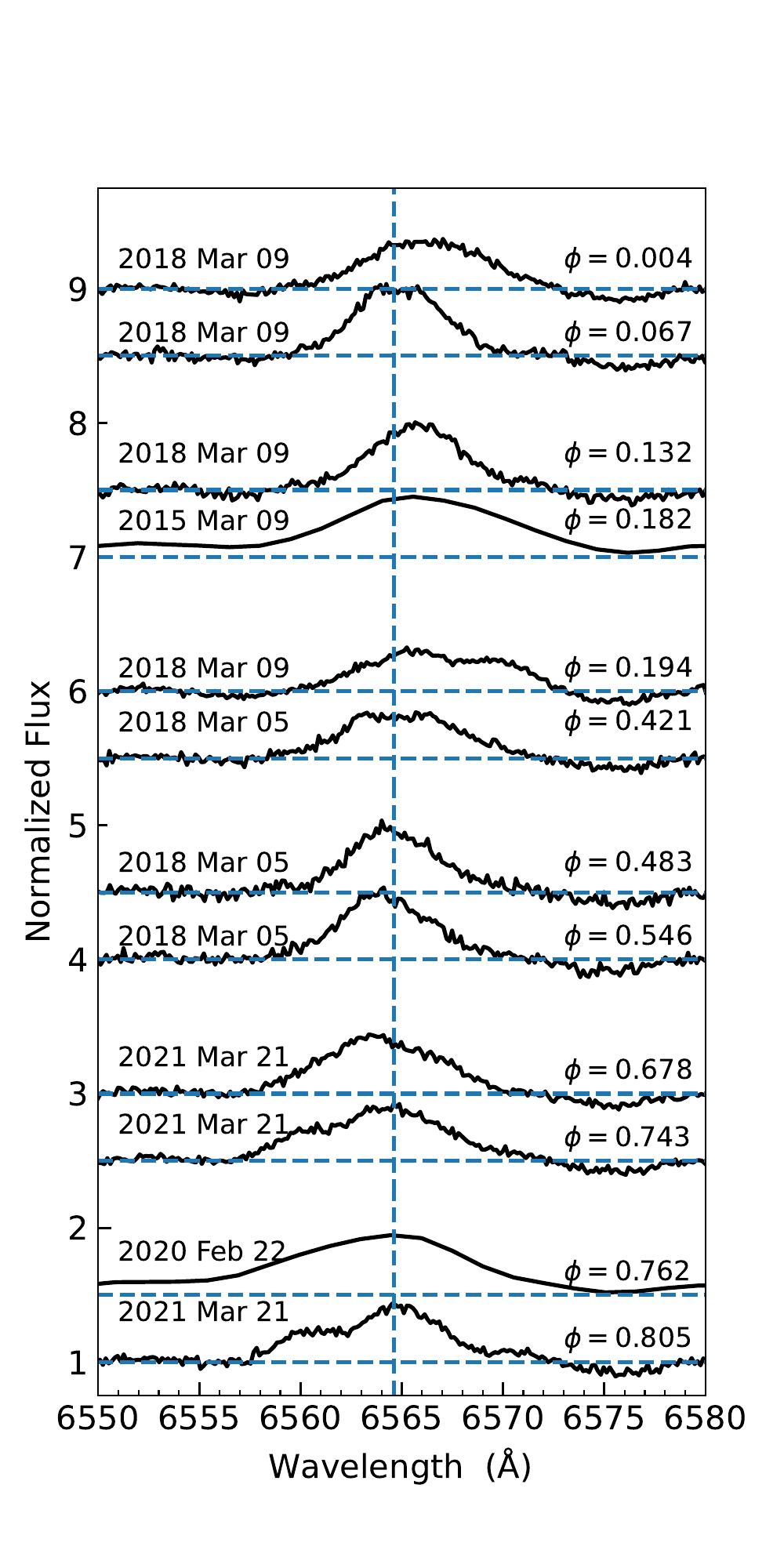}
\includegraphics[angle=0,width=0.49\textwidth]{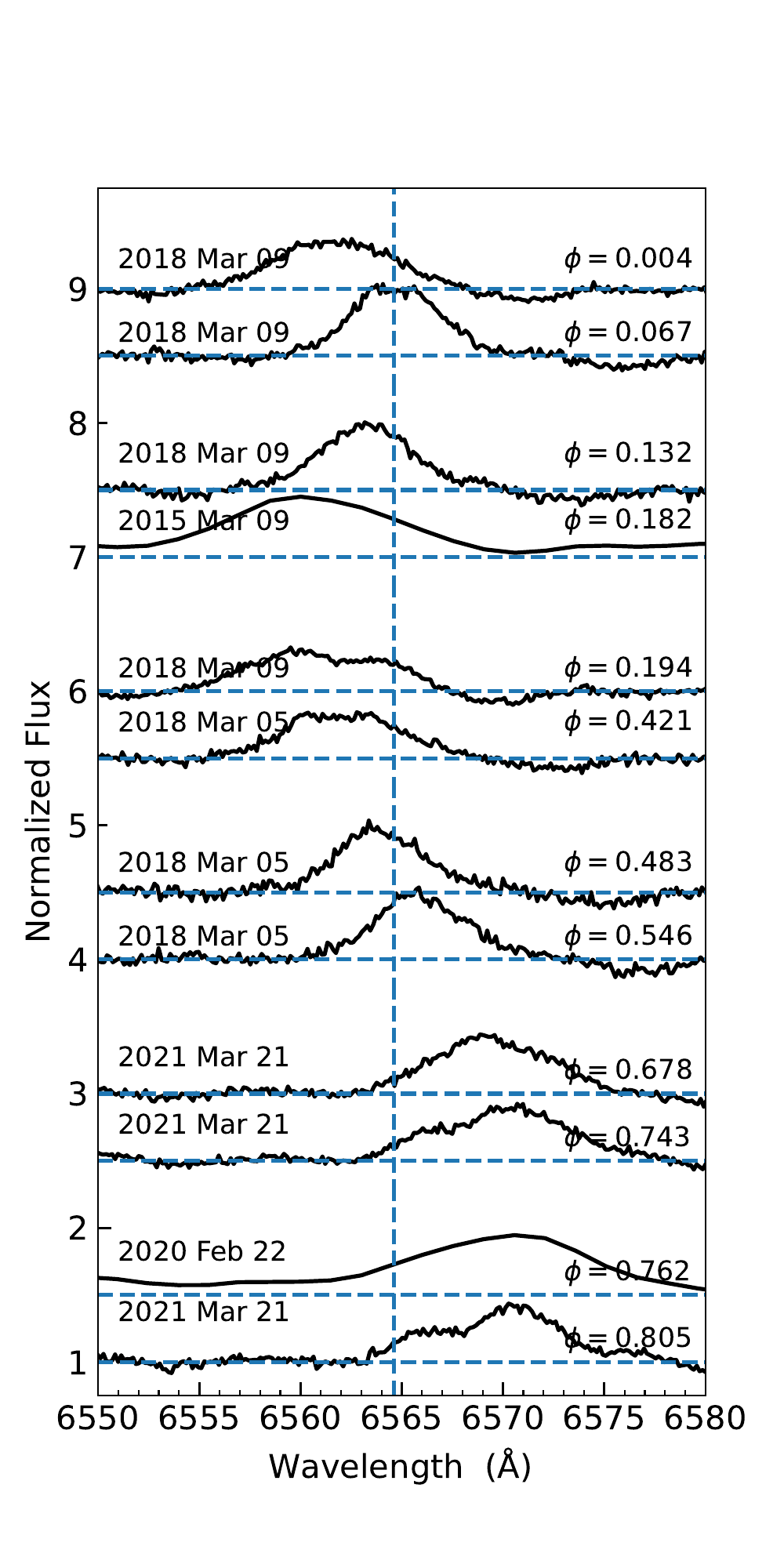}
\caption{Left: $\rm H_{\alpha}$ emission line profiles as a function of orbital phase in the rest frame of the K9-M0 type star.  Right: $\rm H_{\alpha}$ emission-ine profiles in the rest frame of the compact object.}
\label{Halpha}
\end{figure}

\subsection{Radial-velocities curves}
After correcting the observation epochs to the Heliocentric Julian Date (HJD), we performed a circular Keplerian model fit to our radial-velocity data using the custom Markov Chain Monte Carlo sampler TheJoker \citep{2017ApJ...837...20P}. Here, we fit the four free parameters for the period $P$, the time of ascending node $T_{0}$, systemic velocity $\gamma$ and the semiamplitude $K$. We obtain $P=0.2556698\pm 0.0000002$\,days, $T_{0}=2457091.607\pm0.001$\,days, $K=171.09_{-0.97}^{+1.00}$\,$\rm km$\,$\rm s^{-1}$ and $\gamma=-31.89_{-0.80}^{+0.82}$\,$\rm km$\,$\rm s^{-1}$. We find that the spectroscopic period is basically consistent with the photometric period reported by the ASAS and CSS. The phased radial-velocity curve is shown in Figure~5. We used the posterior samples from this fit to derive the mass function $f(M)$
\begin{equation}
	f(M) = \frac{PK^{3}}{2\pi G}= \frac{M_{\rm p}^{3}{\rm sin}^{3}i}{1+q}
\end{equation}
for mass ratio $q=M_{\rm p}/M_{\rm c}$ and inclination $i$. Here, we define the visible main-sequence star as the primary $M_{\rm p}$, while $M_{\rm c}$ is the dark companion. We find $f(M)= 0.131 \pm 0.002~\rm M_{\odot}$. Using the measured rotational velocity and semi-amplitude K, the mass ratio is $q=0.63\pm0.06$ given by the standard equation $v\rm sin$$i$$=0.462Kq^{1/3}(1+q)^{2/3}$ \citep{2001LNP...563..277C}.

\begin{figure}[htp]
\centering
\includegraphics[width=15cm]{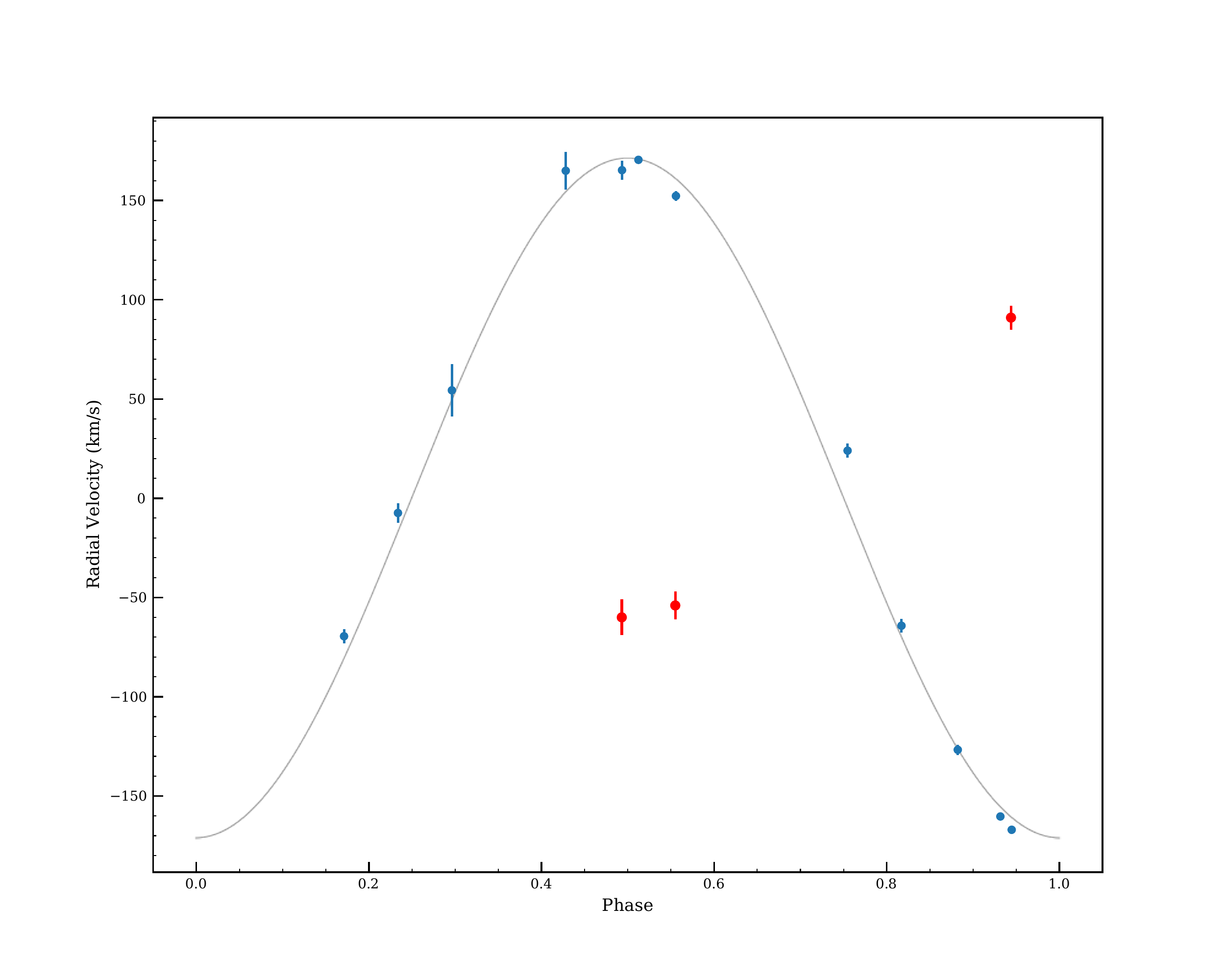}
\caption{The observed radial velocities for the J1527 as a function of the orbital phase obtained in 12 epochs from the LAMOST, with the radial-velocities fit, plotted. The red velocity points are the radial velocities of the narrow $\rm H_{\alpha}$ emission, which is in antiphase with the absorption-line velocities}
\end{figure}

\subsection{Light-curve Modelling }
The light curves of J1527 clearly show a significant ellipsoidal variability with a rather small amplitude. The light curves are asymmetric in the phased light curve of the TESS T-band, showing the fluxes of the two maxima are different, while there are two maxima and one minimum in the phased light curve of the B, V, and R-band. This suggests that the starspot patterns evolve with time as the chromospheric activity changes. Here, we assume that the visible star filling its Roche lobe, given the evidence for an accretion disk, and the accretion disk do not contribute significantly to the optical luminosity of the system due to the low X-ray luminosity of the system. Indeed, there is also no evidence of irradiation in the light curve.

To fit the B, V, and R-band light curves, we had to add one spot on the K9-M0 star surface with the longitude, latitude, angular radius, and temperature factor as their parameters while two spots are needed for the TESS T-band.  We fit the B, V, and R-band light curves and independently fit the TESS T-band using PHOEBE 2.4 \citep{2020ApJS..250...34C}. For the dark companion, we used the option `distortion method = none' and assume a small ($R=3\times10^{-6} \rm~R_{\odot}$), cold ($T_{\rm eff}=300 \rm~K$) blackbody, as done in \cite{2021MNRAS.504.2577J}. We set the gravity darkening cofficient as $\beta=0.32$ and limb-darkening coefficients as (x,y)=(0.5,0.5). We do not include the effects of irradiation. We initially performed trial fits to the B, V, and R-band light curves using Nelder-Mead simplex optimization routine. Then, the parameters from trial fit to performed an MCMC run (nwalkers=48,niters=1000,burnin=150) using the emcee \citep{2019JOSS....4.1864F} solver in PHOEBE\,2.4. We fit over the following parameters: the binary mass ratio $q=M_{\rm p}/M_{\rm c}$, orbital inclination ($i$), the effective temperature ($T_{\rm eff}$) and one spot of the main-sequence K9-M0 star in which the parameters of the starspot include the longitude, latitude, angular radius, and temperature factor\footnote{The ratio of the temperature of the spot to the local intrinsic value.}. The results of the fitting are shown in Table~4.  We also independently fit the TESS T-band light curve using two star spots and fixed the parameters of $q$, $i$, $a$ and $T_{\rm eff,K}$ derived from the previous fitting results. The B, V, R, and T-band light curves with the best-fitting model from PHOEBE are shown in Figure 6, while the residuals of the model are shown in Figure~8 (see Appendix A). The effective temperature and Roche-lobe radius ($ 0.67\pm0.01 \rm R_{\odot}$) of the system derived from the PHOEBE model basically agree well with those obtained from the SED. Combining the modeling of the radial velocity and the PHOEBE model for the ellipsoidal variations, we have enough information to directly determine the masses of the two components in the system. The companion mass is 
\begin{equation}
	M_{\rm c} = \frac{f(M)(1+q)^2}{{\rm sin}^{3}i}
\end{equation}
and from the PHOEBE models, we find that the K9-M0 main-sequence star has a mass $M_{\rm p}=0.62 \pm 0.01~\rm M_{\odot}$ and the companion mass is $M_{\rm c}=0.98 \pm 0.03~\rm M_{\odot}$. Based on the mass-luminosity relations from \cite{2019ApJ...871...63M}, we obtained that the visible star has a mass of $0.63\pm0.02 \rm M_{\odot}$, which is consistent with the one from the PHOEBE models. For the Roche lobe filling factor $<1$, we obtained a K9-M0 main-sequence star of $M_{\rm p}=0.52\pm 0.06~\rm M_{\odot}$ and an unseen companion of $M_{\rm c}=0.87 \pm 0.10~\rm M_{\odot}$. Here, the orbital inclination and mass ratio $46.65^\circ{}^{{+0.59}^{\circ}}_{-0.68^{\circ}}$ and $0.596^{+0.041}_{-0.065}$, respectively. However, the obtained mass and radius ($ 0.620^{+0.014}_{-0.035} \rm R_{\odot}$) of the visible star are small than those resulting from the mass-luminosity relations and SED, which may suggest that the visible star is filling its Roche lobe (Table 1).

\begin{table}
\begin{center}
\caption{The best-fitting parameter for the B, V, R-band and T-band in the filling visible Roche lobe, respectively.}\label{tab:params}
	\begin{tabular}{r c c c }
	\hline	
	& Parameter & B, V, and R-band   &   T-band \\
	\hline
	\vspace{2mm}
	& $P_{\rm orb}$  (days)           & \multicolumn{2}{c}{0.2556698 (fixed) } \\
	\vspace{2mm}
	& $a~(R_{\odot})$              & \multicolumn{2}{c}{$1.971^{+0.019}_{-0.011} $ }\\
	\vspace{2mm}	
	& $i~(^\circ)$                 & \multicolumn{2}{c}{$45.20^{+0.13}_{-0.20} $ }\\
	\vspace{2mm}	
	& $T_{\rm{eff}}~(K)$           & \multicolumn{2}{c}{$3894.95._{-0.92}^{+0.69} $} \\
	\vspace{2mm}
	& $q$                          & \multicolumn{2}{c}{$0.631_{-0.003}^{+0.014} $} \\
	\vspace{2mm}
        &Spot 1 longitude ($^{\circ}$) & $183.80 ^{+0.28}_{-0.28}$     & $315.12^{+0.29}_{-0.66}$ \\
	\vspace{2mm}
        & Spot 1 colatitude ($^{\circ}$) & $78.83 ^{+0.80}_{-0.35}$      & $65.32^{+0.72}_{-1.85}$ \\
	\vspace{2mm}
        &Spot 1 radius ($^{\circ}$)    & $47.94 ^{+0.25}_{-0.37}$      & $52.04^{+0.51}_{-4.8}$ \\
	\vspace{2mm}
        &Spot 1 temp. factor          & $0.886 ^{+0.001}_{-0.003}$  & $0.922^{+0.001}_{-0.012}$ \\
	\vspace{2mm}
        &Spot 2 longitude ($^{\circ}$) & $-$                           & $157.57^{+0.65}_{-0.43}$ \\
	\vspace{2mm}
        &Spot 2 colatitude ($^{\circ}$)  & $-$                           & $81.95^{+0.49}_{-1.82}$ \\
	\vspace{2mm}
        &Spot 2 radius ($^{\circ}$)    & $-$                           & $24.07^{+2.21}_{-0.30}$ \\
	\vspace{2mm}
        &Spot 2 temp. factor           & $-$                           & $0.691^{+0.071}_{-0.013}$ \\
	\hline
	\vspace{2mm}	
\end{tabular}
\end{center}
\end{table}

\begin{figure}[htp]
\centering
\includegraphics[width=15cm]{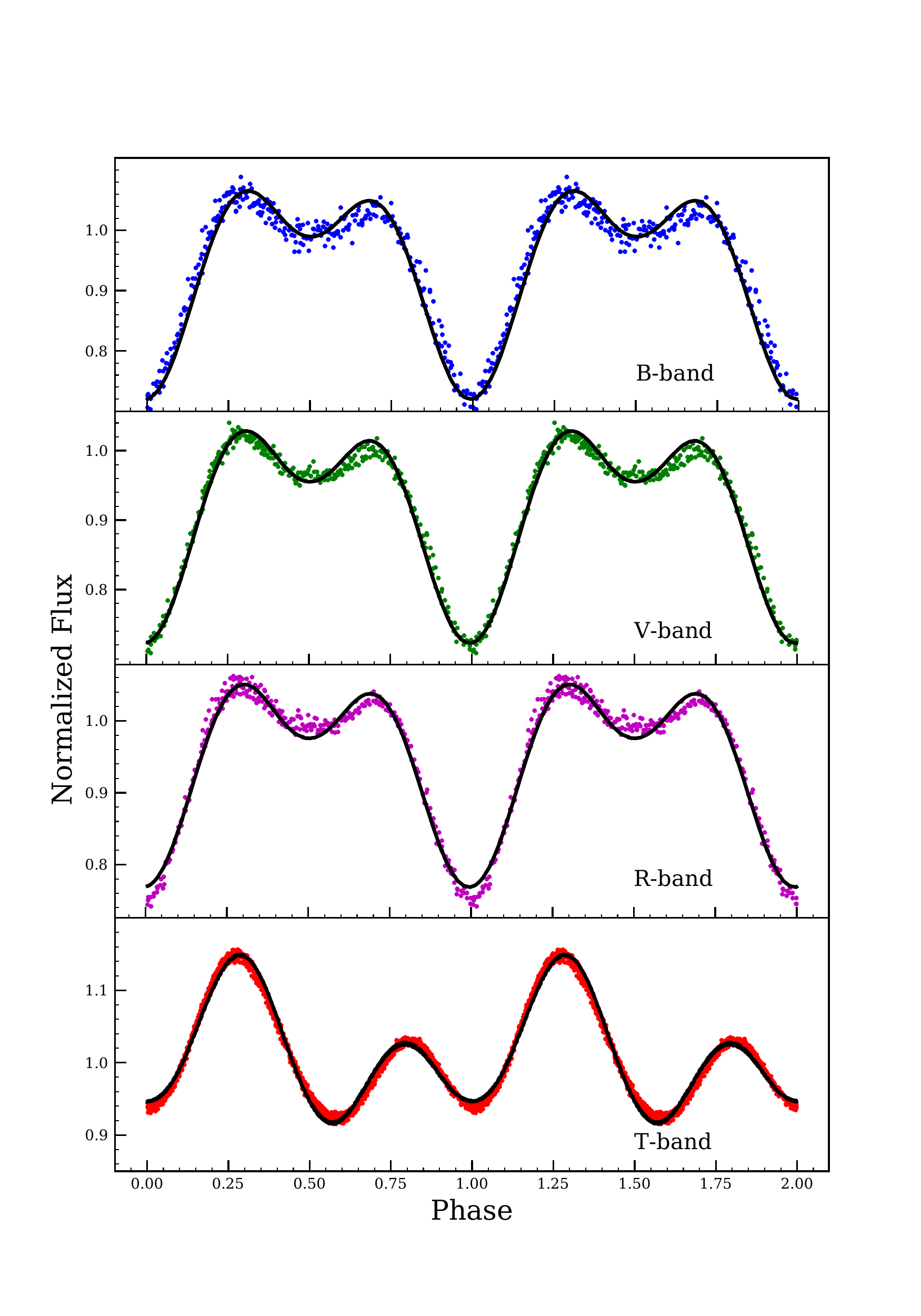}
\caption{The normalized B, V, R-band and TESS light curves for  the J1527 as a function of orbital phase. The black solid line represents the best-fitting model from PHOEBE.}
\end{figure}

\subsection{Radio and X-Ray detection}
The FAST searches have failed to identify a radio pulsar signal from the J1527. In our 40 minute and 50 minute searches, we can infer a flux density upper limit of 6 $\mu$Jy and 5 $\mu$Jy at 1250 MHz, respectively. Here, we assume a minimum detectable signal-to-noise ratio $S/N_{min}=10$ and a 20 percent duty cycle. At the Gaia EDR3 distance, we obtain the X-ray luminosities of (9.30 $\pm$ 3.32) $\times10^{28}$$\rm~erg$\,$\rm s^{-1}$ in the 0.1-2.4\,keV PSPC range of the RASS. The low X-ray luminosity of the system is consistent with a quiescent dwarf nova or the chromospherically active RS CVn system \citep{1987Ap&SS.137..195D}. Therefore, the X-ray emission can originate from one of them or a mixture of both but may also an NS.


\section{Discussion}
\subsection{Is the compact companion in J1527 a white dwarf?}
The mass of the compact object ($M_{\rm c}=0.98 \pm 0.03~M_{\odot}$) in J1527 may suggest it can be a white dwarf in a CV (an interacting binary system with a white dwarf companion). However, we lay out the arguments that the compact companion in J1527 as a white dwarf will suffer from some trouble. It is found that the narrow component of H$_{\alpha}$ is in antiphase with the
radial-velocity curve of the visible star, which may give an indication of the possible presence of an accretion disk. The result helps us immediately rule out that the source belongs to the polar class. This is because the polars have a strong magnetic field in the range of $\sim$10-300\,MG, preventing the formation of an accretion disc \citep{1990SSRv...54..195C}. For a polar-type CV, it is worth noting that a broad H$_{\alpha}$ component may also originate in an accretion stream between the two stars \citep{2020A&A...637A..35S}. However, the narrow component originates from the unseen companion in J1527, while the broad component is in phase with the absorption-line radial velocity. If the narrow component is due to an accretion stream rather than an accretion disk, the compact star may be a polar. In the intermediate polars (IPs), the magnetic field is somewhat smaller and a partial accretion disk is usually present, and the white dwarf does not corotate with the orbit \citep{1994PASP..106..209P}. Here, the source has a very low X-ray luminosity, which is not consistent with any known X-ray luminosity of the IP. In addition, the observed low X-ray luminosity is also consistent with low-accretion rate polars (LARPS) when a white dwarf accretes at an extremely low rate from the wind of the visible donor star \citep{2009A&A...500..867S}. Therefore, the source as a candidate for a polar or an IP looks problematic if the narrow H$_{\alpha}$ component indeed is from an accretion disk. However, it is not possible to assess whether it is a polar or an IP using the present data.                  

A final discussion concerns the dwarf novae, a class of non-magnetic CVs in which the system undergoes outbursts to brighten by several magnitudes lasting from days to weeks. The outbursts are thought to be caused by disk instabilities when the accretion disk reaches a critical density. The J1527 has been detected by the deep GALEX far- and near-ultraviolet (FUV and NUV), and the measured flux of the NUV is about five times higher than that of the FUV, which indicates an ultraviolet excess from the K9-M0-type star. In fact, the observed \ion{Ca}{2}\,H\&K and $\rm H_{\alpha}$ emission are present in the low-resolution spectra of J1527 as the common indicators of chromospheric activities, which seemingly suggest that the deep GALEX FUV and NUV result from stellar chromospheric activity. However, the luminosity of the FUV (4.07 $\pm$ 0.57 $\times10^{29}$$\rm~erg$\,$\rm s^{-1}$) and NUV (2.10 $\pm$ 0.09 $\times10^{30}$$\rm~erg$\,$\rm s^{-1}$) is an order of magnitude higher than the chromospheric luminosity, which is not consistent with the chromospheric activities of the late-type star \citep{2016MNRAS.463.1844S}. However, we do not exclude the chromospheric contribution from the late-type star. In order to subtract the UV contribution from chromospheric activity, we estimate the chromospheric contribution at FUV and NUV. The median EW of the $\rm H_{\alpha}$ emission is $\sim$ 2.72. We can convert the $\rm H_{\alpha}$ at the surface using $ F_{\alpha}=\rm EW(\alpha)$$F_{\rm c}$ \citep{1993ApJS...85..315S}, where $F_{\rm c}$ is the continuum flux at $\rm H_{\alpha}$ that we drive using the flux calibration of \cite{1996PASP..108..313H}, log$F_{c}=7.538-1.08(B-V)_{0}$. We obtain $(B-V)_{0}=1.28\pm0.01$ mag using the APASS DR10 photometry. Then, using the empirical relation (log$L_{\rm NUV}/L_{\rm bol}=0.67\rm log(L_{\alpha}/L_{\rm bol})-0.85$) of \cite{2016ApJ...817....1J}, we can obtain the NUV of $\sim$ $1.24\times10^{-16}$ $\rm erg \rm s^{-1} \rm cm^{-2} \mathring {A}^{-1}$. For FUV, we obtain the $R'_{\rm HK}=4.1$ using the log$R'_{\rm HK}$-log$P_{\rm rot}$ relationship of \cite{2017A&A...600A..13A}. According to the relation (log$R'_{\rm FUV}$=(0.98$\pm$0.05)log$R'_{\rm HK}$+(-0.53$\pm$0.25)) of \cite{2011AJ....142...23F}, the FUV flux can be calculated as $\sim$ $1.46\times10^{-17}$ $\rm erg \rm s^{-1} \rm cm^{-2} \mathring {A}^{-1}$. Here, we subtract the visible-star flux and the chromospheric contribution at FUV and NUV.

Therefore, the FUV and NUV excess may stem from either a white dwarf or an accretion disk. if we assume that the FUV excess of the J1527 is from the white dwarf, its effective temperature $T_{\rm eff}$ is about $12000$\,K by adopting a mass of $M\rm_{WD}=1.0~M_{\odot}$ and a radius of $R \sim 5\times10^{8} \rm cm$ (Figure~7). Then, we can utilize the $T_{\rm eff} = 12000$\,K to estimate the accretion rates $\dot{M}$ using the equation of \cite{2009ApJ...693.1007T}:
\begin{equation}
     T_{\rm eff} = 17000 (<\dot{M} [\rm M_{\odot} \rm yr^{-1}]>/10^{-10})^{1/4} (M_{\rm wd} [\rm M_{\odot}]/0.9)~K
\end{equation}. 
Using the estimated temperature of the white dwarf derived from the FUV emission, we can obtain accretion rates of $\dot{M} \sim 1.63\times10^{-11}$\,$\rm M_{\odot}$\,$\rm yr^{-1}$.
The inferred value of the mass-transfer rate is far below the critical accretion rate that would be needed to keep the thermal-viscous disk in a stable state at the this orbital period \citep{2018A&A...617A..26D}, which suggests that the J1527 should exhibit dwarf nova phenomena. However, there is no evidence for the dwarf nova outburst in our long time-series archives data (see Figure~1). We also note that the fraction of dwarf novae above the period gap can remain in quiescence due to the low accretion rates or large accretion disk \citep{2022ApJ...934..142S}. If we consider that approximately half of the gravitational energy of the accreting gas is liberated through X-rays in the boundary layer, then for an $M_{\rm c}\sim 1.00 ~M_{\odot}$ white dwarf, the inferred accretion rates of $\dot{M} \sim 1.63 \times 10^{-11}$\,$\rm M_{\odot}\rm yr^{-1}$ implies the X-ray luminosity of $\sim$ 1.38 $\times10^{32}$\,$\rm erg$\,$\rm s^{-1}$. If assume that the observed X-ray luminosity originates entirely from the released gravitational energy of the accreting gas, it is a factor of $\sim$ 1000 below the inferred X-ray luminosity. However, if white dwarf has a lower temperature in J1527, it will also mean a lower accretion rate or X-ray luminosity. In summary, we cannot completely rule out that the compact object is a white dwarf in J1527. 

\begin{figure}[htp]
\centering
\includegraphics[width=15cm]{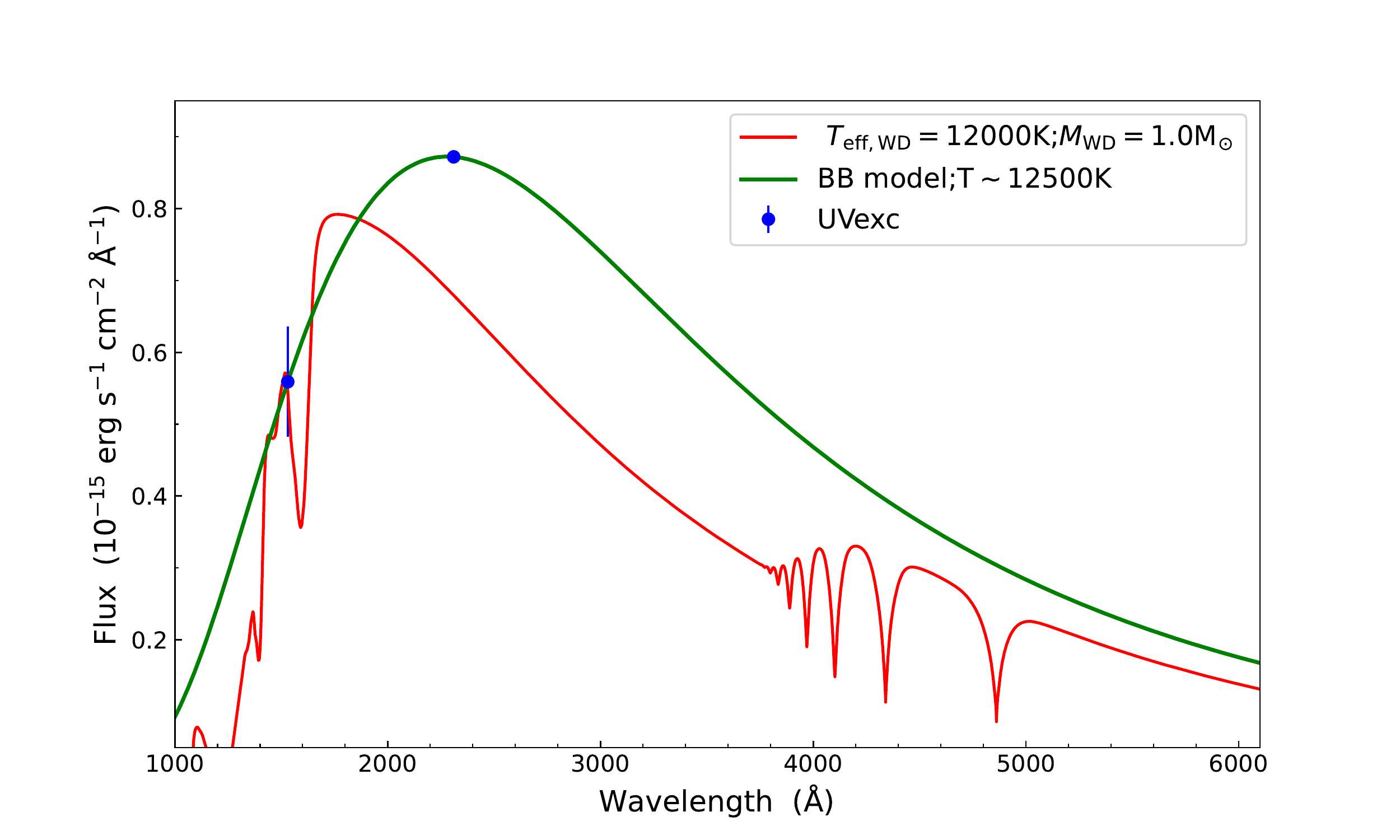}
\caption{The green line is a blackbody model, while the red line is a DA-type white dwarf of $1.0 \rm M_{\odot}$ with $T_{\rm eff,WD}=12000 \rm K$.}
\end{figure}

\subsection{Is the compact companion an ``X-Ray-dim isolated neutron star'' in binary J1527?}
Then, we consider a scenario where the compact companion is a low-mass NS.  
If the compact companion is not a white dwarf in J1527, then it is an low-mass X-ray binary (LMXB) containing either an NS or BH companion. However, the mass of the compact companion immediately rules out a stellar BH, which means the most likely scenario is an NS.
We did not detect the radio pulsations from the J1527 in two observations using FAST, which may suffer from severe scattering and/or absorption due to the system being enshrouded by intra-binary material. In addition, the lack of the detection of radio pulsations in two observations may result from eclipses due to the ephemeris are probably not being accurate enough. If so, there should be a $\gamma$ ray counterpart due to the interaction between the pulsar wind and the accretion disk. However, we did not find a possible
$\gamma$ ray counterpart to J1527 from the Fermi-LAT catalog. Therefore, we propose that the dark
companion of J1527 is likely an XDINS-like compact object in binary.

The most direct interpretation to
account for a low X-ray luminosity is that none or parts of the accretion flow reach the NS surface when the magnetospheric radius $r_{\rm m}$ is much larger than the corotation radius $r_{\rm co}$ according to standard accretion theory \citep{1975A&A....39..185I}. Here, we interpret that the FUV and NUV come from an accretion disk \citep{2014ApJ...785..131T}, which is consistent with the presence of a narrow H$_{\alpha}$ emission. We use one blackbody to fit the FUV and NUV with the Python package of lmfit \footnote{https://lmfit.github.io/lmfit-py/}. The modeled temperature is $T = 12500^{+500}_{-690}$\,K with the corresponding radius $R = 5.23^{+0.82}_{-0.48} \times 10^{8}$\,cm. Here, we make the simplified assumption that the corresponding radii $R_{\rm BB}$ is consistent with the magnetospheric radius $r_{\rm m}$. The magnetospheric radius is 
\begin{equation}
  r_{\rm m}=1.2\times10^{9}\mu^{4/7}_{29}\dot{M}_{13}^{-2/7} (\frac{M}{M_{\odot}})^{-1/7} \sim 5.23\times 10^{8} \rm~cm,
\end{equation}
where the magnetic moment $\mu=B_{\rm p}R^{3}/2\sim10^{29}B_{\rm p,11}~\rm G$$\rm cm^3$ and the radius of the NS is $R=12 \rm~km$. The observed $L_{\rm bol}=GM\dot{M}/2r_{\rm m} \sim 4.78 \times 10^{30}$\,$\rm erg$$\rm s^{-1}$ derived from the one-blackbody model, which gives the disk accretion rate $\dot{M}=3.74 \times 10^{13}$\,g\,$\rm s^{-1}$. According to Equations (4) and the disk accretion rate, we obtain a weak magnetic field of $\sim B_{\rm P}=4.53 \times 10^{10}~\rm G$ for an NS. The system is in the propeller regime when the magnetospheric radius is larger than the corotation radius ($r_{\rm m} > r_{\rm c}$). In this picture, the NS with a weak magnetic field must have a spin period $P_{\rm spin} < 6.50 \rm s $, since the corotation radius is 
\begin{equation}
  r_{c}=(GMP^{2}/4\pi^{2})^{1/3}=1.50\times10^{8}P^{2/3} \rm~cm.
\end{equation}
Considering a radio-quiet NS with a weak magnetic field, it may be located at the radio pulsar death line because the electric potential of the gap region is too low to generate electron-positron pairs. This also accounts for the absence of the $\gamma$ ray in the system. Using the radio pulsar death line $B_{\rm p}/P^{2}=1.7\times10^{11}$\,G\,$\rm s^{-2}$ \citep{1992A&A...254..198B}, we obtain the spin period of death line $P_{\rm spin}\sim 0.52$\,s. Therefore, we estimate the spin period of the neutron star $0.52 \rm s \leq P_{\rm spin}\leq 6.50$\,s, which also is consistent with the characteristic spin period of the XDINS \citep{Kaplan11}.  In addition, the hardness ratio from the RASS uncovers a relatively soft emission. This may suggest that the X-ray emission is from an NS although the soft X-ray emission also appears to be consistent with a chromospheric origin.

In addition, a dynamically discovered NS with $M_{\rm c}\sim 1.00$\,$M_{\odot}$ will be the lightest neutron star, which challenges the paradigm of gravitational-collapse neutron star formation \citep{2012ARNPS..62..485L}. However, a low-mass limit ($0.9-1.1$\,$M_{\odot}$) is also suggested when considering that both thermal and neutrino-trapping effects are large \citep{1998A&A...330.1005G,1999A&A...350..497S}. If supernova explosions cannot possibly produce an NS with a mass smaller than about $1.17M_{\odot}$ \cite{2018MNRAS.481.3305S}, the compact object of J1527 may be a strange star. Recently, the analysis of the central compact object (CCO) within the supernova remnant HESS J1731-347  reported an NS with $M=0.77^{+0.20}_{-0.17} M_{\odot}$ based on modeling of the X-ray spectrum and a robust distance \citep{2022NatAs...6.1444D}. In addition, \cite{2016MNRAS.458.2565D} present evidence that the CCO of the supernova remnant HESS J1731-347 could have been formed within a binary system. Based on some similarities between the neutron star of both HESS J1731-347 and J1527, this may imply that they may be born via the same channel. More interestingly, the discovery that the radionuclide $^{60} \rm Fe$ signal observed in deep-sea crusts is global indicates multiple supernova events during the last 10 million years within $\sim 100$\,pc of Earth \citep{2016Natur.532...69W,2022arXiv220606464E,2022arXiv221004685Z}. The XDINS-like compact object at a distance of $\sim 118$\,pc makes it the nearest neutron star, which may suggest the radionuclide $^{60} \rm Fe$ signal from the deep-sea crusts is associated with its supernova event. Furthermore, we performed a kinematic analysis of J1527's orbit in the galaxy using the Gaia astrometric solution and the systematic radial velocity (See Appendix B), and it implies that it can pass through our solar neighborhood and is consistent with residing in the Galactic thin disk.

The discovery of a compact object in J1527 likely implies that some XDINS-like objects remain in a binary when they receive low kicks at birth. As a result, there may be lots of XDINS-like stars, either as a single or in binaries, in the Milky Way since that these NSs are dim and close to us. Our findings may hint at XDINS-like compact objects would be born in an alternative channel rather than in standard core-collapse supernovae, as discussed for the accretion-induced collapse (AIC) of an ONeMg white dwarf \citep{2022arXiv221008125T}, for instance.



\section{Conclusions}
We identify a K9-M0-type star with chromospheric activity in a binary system with a dark companion mass of $M_{\rm c}=0.98 \pm 0.03$\, $\rm M_{\odot}$. By modeling the multiband light curve with PHOEBE, we derive an inclination of $45.20^\circ{}^{+0.13^{\circ}}_{-0.20^{\circ}}$, a mass ratio of $0.631^{+0.014}_{-0.003}$, and a  K9-M0-type star's mass of $0.62 \pm 0.01$\,\rm $M_{\odot}$ using constraints on the radial velocity, orbital period, and stellar temperature. The medium-resolution LAMOST spectrum uncovers the $\rm H_{\alpha}$ emission of wider range and another peak emission, suggesting the presence of an accretion disk. Here, the dark companion mass of $M_{\rm c}=0.98 \pm 0.03$\,$M_{\odot}$ is either a white dwarf or an NS. However, the system shows some characteristics that seem to contradict it being a cataclysmic variable, such as the lack of any observed outburst in the 5193\,d time-series of optical observations. Therefore, we discuss the other possibility that the unseen compact companion is an low-mass NS. If the dark companion is confirmed as an NS, it will be nearest and lightest NS yet. 
These features, together with being X-ray dim and radio-quiet, are similar to those of XDINSs, and we thus suggest the J1527 binary may host an XDINS-like compact object. 

To determine the nature of the compact object and understand this unique system, further multiband observation are necessary. The observation of the Hubble Space Telescope spectroscopy will Especially determine whether the compact object is a white dwarf, and the X-ray observations may also address the nature of the compact object.

\acknowledgments

We appreciate Jianning Fu, Kejia Lee, Song Wang, Jifeng Liu, Xinlin Zhao, Bojun Wang, Chunyang Cao and Jianping Xiong for their helpful comments and suggestions. The Guoshoujing Telescope (the Large Sky Area Multi-Object Fiber Spectroscopic Telescope, LAMOST) is a National Major Scientific Project built by the Chinese Academy of Sciences. Funding for the project has been provided by the National Development and Reform Commission. LAMOST is operated and managed by the National Astronomical Observatories, Chinese Academy of Sciences. FAST is a Chinese national mega-science facility, operated by the National Astronomical Observatories, Chinese Academy of Sciences. This paper makes use of data from the first public release of the WASP data \citep{2010A&A...520L..10B} as provided by the WASP consortium and services at the NASA Exoplanet Archive, which is operated by the California Institute of Technology, under contract with the National Aeronautics and Space Administration under the Exoplanet Exploration Program. This paper includes data collected with the TESS mission, obtained from the MAST data archive at the Space Telescope Science Institute (STScI). Funding for the TESS mission is provided by the NASA Explorer Program. STScI is operated by the Association of Universities for Research in Astronomy, Inc., under NASA contract NAS 5-26555. We acknowledge the use of the public data from the ASAS, CRST, and HATNet.  This research has made use of the SIMBAD database, operated at CDS, Strasbourg, France.
This work was supported by the National SKA Program of China (2020SKA0120100), the National Natural Science Foundation of China (12090040, 12090044,11833006), and the Strategic Priority Research Program of CAS (XDB23010200).

\appendix
\section{SHORT TIMESCALE VARIABILITY IN THE RESIDUALS}
\label{a}
\begin{figure}[htp]
         \centering
         \includegraphics[width=15cm]{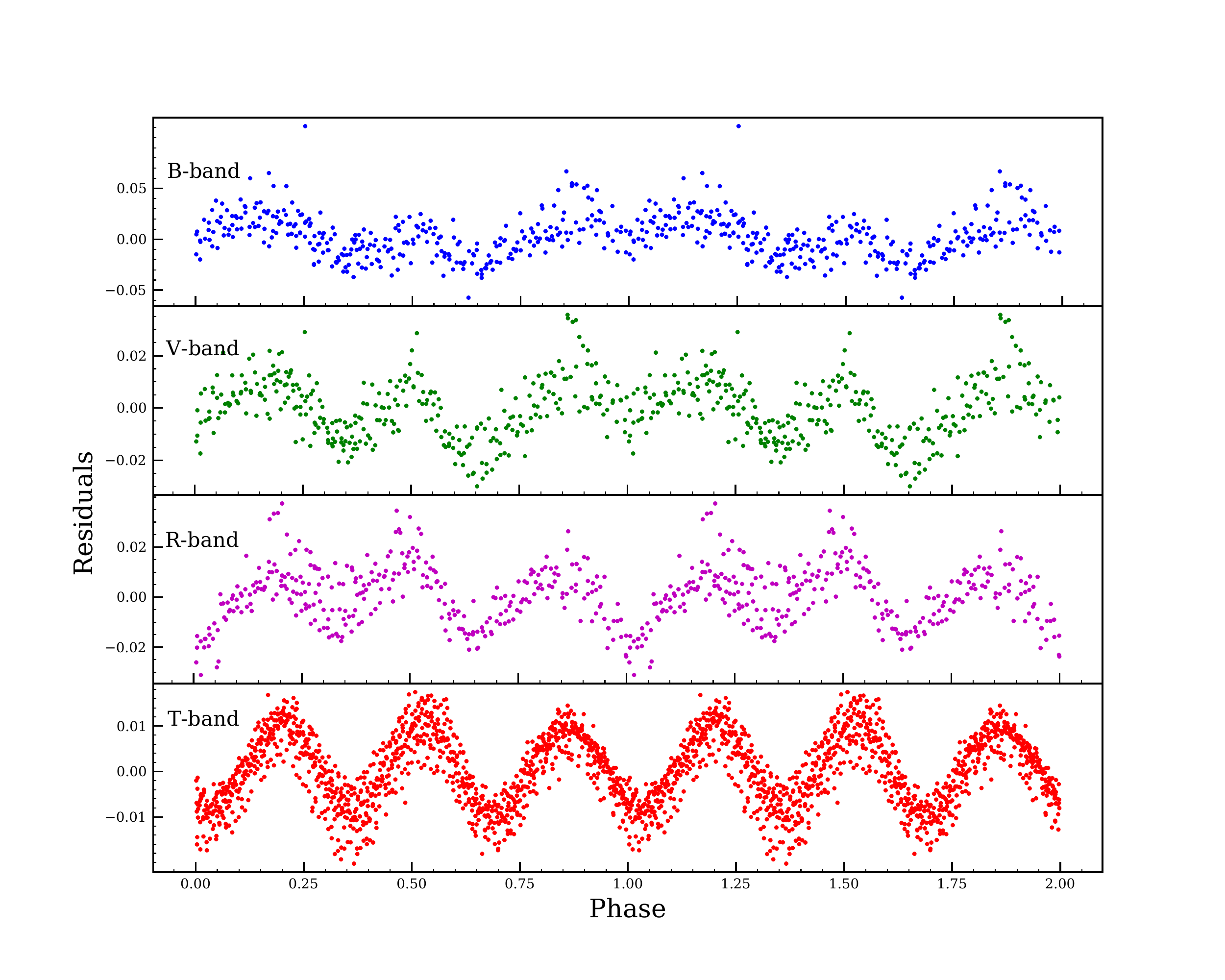}
         \caption{The light curve residuals for multi-band (B,V,R and T) after the PHOEBE model is subtracted.}
         \label{res}
\end{figure}

There are extra variability on the time-scales of hours in the residuals of the B, V,and R bands and TESS T-band light curves after subtracting the best model, which may result from the disk oscillations known as superhumps \citep{2002MNRAS.333..791Z,2017MNRAS.469..950R}. In addition, the residuals exhibit similar characteristics in the light curves (Fig.~\ref{res}). We searched for additional variability on the time-scale of hours in the TESS T-band light curve and find no evidence of variability on any other period. Further, there is no periodic signal after prewhitening the data on the main period using higher-cadence observation for the B, V, and R band. Therefore, the residuals of the light curve show quasi-sinusoidal signals, which might not be induced by the disk oscillations (superhumps). Here, we can see clear evidence of structure in the residuals, which may be from the stellar pulsation, random noise, and systematic calibration noise \citep{2011ApJS..197....4W}. 

\section{The kinematic analysis of J1527}
\label{b}
\begin{figure}
   \centering
   \includegraphics[width=10cm]{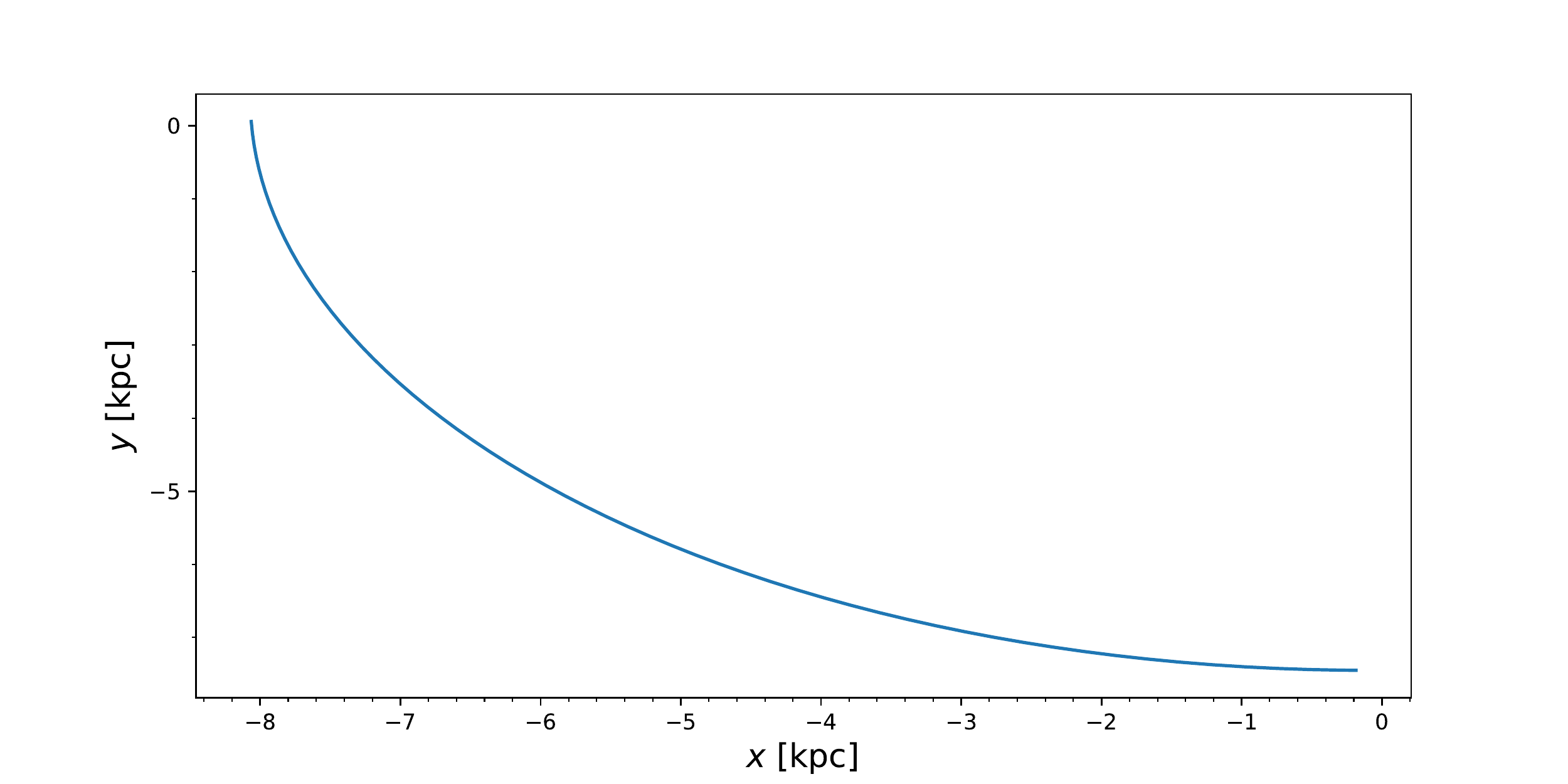}
   \includegraphics[width=10cm]{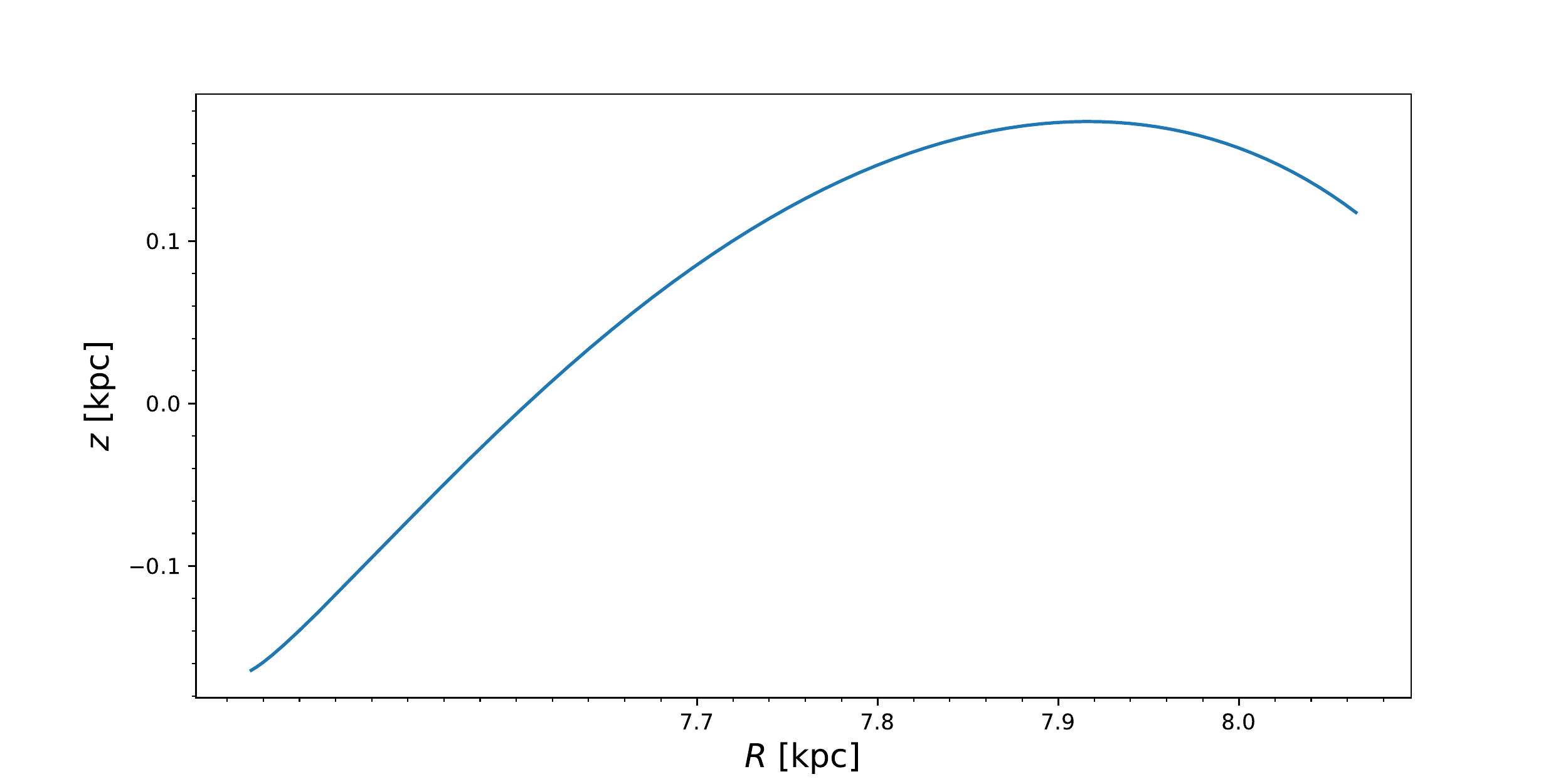}
   \includegraphics[width=10cm]{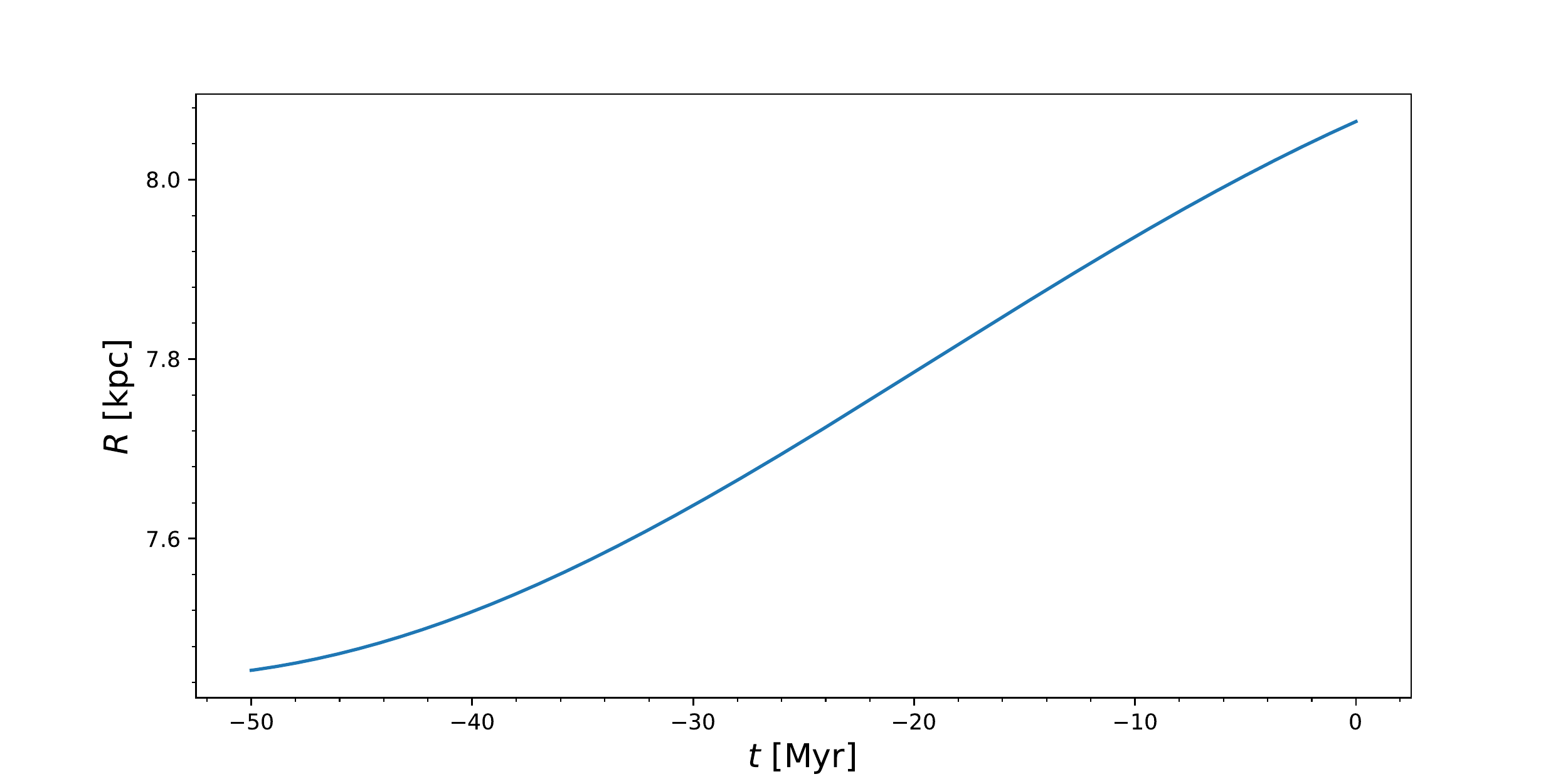}
   \caption{A set of panels shows the trajectory of J1527 around the Milky Way integrated backwards from present day for 50 Myr.}
   \label{xyz}
\end{figure}

We use the Gala code to compute J1527's trajectory around the Milky Way over 50 Myr \citep{2017JOSS....2..388P}, using the Milky Way potential. Here, we use the parallax and proper motion reported by the Gaia, and the systematic radial velocity fitted by our radial-velocity data (Fig.~\ref{xyz} ). This suggests that the J1527 resides in the Galactic thin disk and the system passed through our solar neighborhood in the past. 

\end{document}